\documentclass[twocolumn,twoside,slac_two]{revtex4}
\usepackage{graphicx}
\usepackage{fancyhdr}
\begin{document}
\title{VLBA monitoring of Mrk 421 at 15 and 24 GHz during 2011}

\author{R. Lico, M. Orienti, G. Giovannini}
\affiliation{Dipartimento di Astronomia, Universit\`a di Bologna, 40127 Bologna, Italy}

\author{M. Giroletti}
\affiliation{INAF Istituto di Radioastronomia, 40129 Bologna, Italy}

\author{B. Cotton}
\affiliation{National Radio Astronomy Observatory, Charlottesville, VA 22903-2475, USA}

\author{P. G. Edwards}
\affiliation{CSIRO Australia Telescope National Facility, Marsfield, NSW 2122, Australia}

\author{L. Fuhrmann, T. P. Krichbaum} 
\affiliation{Max-Planck-Institut f\"ur Radioastronomie, Auf dem H\"ugel 69, 53121 Bonn, Germany}

\author{K. Sokolovsky, Y. Kovalev}
\affiliation{Astro Space Center of the Lebedev Physical Institute, 117997 Moscow, Russia}

\author{S. Jorstad, A. Marscher}
\affiliation{Institute for Astrophysical Research, Boston University, Boston, MA 02215, USA}

\author{M. Kino}
\affiliation{National Astronomical Observatory of Japan, Osawa 2-21-1, Mitaka, Tokyo 181-8588}

\author{D. Paneque}
\affiliation{Max-Planck-Institut f\"ur Physik, D-80805 M\"unchen, Germany}

\author{M. A. Perez-Torres}
\affiliation{Instituto de Astrofísica de Andalucia, IAA-CSIC, 18080 Granada, Spain}

\author{G. Piner}
\affiliation{Department of Physics and Astronomy, Whittier College, Whittier, CA, USA}

\begin{abstract}
We present a preliminary analysis of new high resolution radio observations of 
the nearby TeV blazar Markarian 421 (z=0.031). This study is part of an
ambitious multifrequency 
campaign, with observations in sub-mm (SMA), optical/IR (GASP), UV/X-ray (Swift,
RXTE, MAXI), 
and $\gamma$ rays (Fermi-LAT, MAGIC, VERITAS). In this manuscript we consider only data obtained 
with the Very Long Baseline Array (VLBA) at seven epochs (one observation per
month 
from January to July 2011) at 15 and 23.8 GHz. We investigate the inner jet
structure on parsec scales through the study of model-fit components for each epoch.
We identified 5-6 components which are consistent with being stationary during the 6-month period reported here. 
The aim is to try to shed light
on questions 
such as the nature of radiating particles, the connection between radio and
$\gamma$-ray emission, 
the location of emitting regions and the origin of the flux variability.
\end{abstract}

\maketitle

\thispagestyle{fancy}

\begin{table*}
\begin{center}
\footnotesize
\begin{tabular}{cccccccc}
\hline
\textbf{Observation} &  \multicolumn{2}{c}{\textbf{Map Peak}} & 
\multicolumn{2}{c}{\textbf{Beam}} & \multicolumn{2}{c}{\textbf{Lowest contour}} 
& \textbf{Notes}  \\
\textbf{date}  &  \multicolumn{2}{c}{\textbf{(mJy/beam)}} & 
\multicolumn{2}{c}{\textbf{(mas x mas)}} &
\multicolumn{2}{c}{\textbf{(mJy/beam)}}  &  \\
      & 15GHz & 23.8GHz & 15GHz & 23.8GHz& 15GHz & 23.8GHz&   \\
\hline
\hline
2011/01/14   & 340 & 316 &\ \ 0.89 x 0.52  \ \ &\ \ 0.73 x 0.41 \ \ &
1.0 & 1.0 & No MK, no NL    \\
2011/02/25  & 391 & 335 & 1.09 x 0.70  & 0.58 x 0.35 & 0.8 & 0.7 & NL
snowing \\
2011/03/29 & 384 & 358 & 0.93 x 0.55 & 0.61 x 0.36 & 1.2 & 1.1 & No HK 
\\
2011/04/25  & 358 & 305 & 0.89 x 0.50 & 0.56 x 0.32 & 1.1 & 0.9 & -  \\
2011/05/31   & 355 & 295 & 0.90 x 0.52 & 0.58 x 0.34 & 1.1 & 0.9 & - 
\\
2011/06/29  & 260 & 207 & 0.85 x 0.49 & 0.56 x 0.32 & 1.0 & 0.8 & No LA
  \\
2011/07/28  & 228 & 192 & 0.87 x 0.51 & 0.52 x 0.34 & 0.9 & 0.8 & -  
\\
\hline
\end{tabular}
\caption{Details of observations.\label{observations}}
\end{center}
\end{table*}

\section{Introduction}
Markarian 421 (R.A.=$11^h04^m27.31^s$, decl.=$+38^\circ$ $12'$ $31.8''$) is one
of the nearest (z=0.031) and 
one of the brightest BL Lac objects in the sky. It was the first extragalactic
source detected at 
TeV energies by the Cherenkov telescope at Whipple Observatory \cite{Punch1992}. The
spectral energy 
distribution (SED) of this object, dominated by non-thermal emission, shows two
smooth broad components: 
one at lower energies, from radio band to soft X-ray domain, and one at higher
energies peaking at $\gamma$-ray energies \cite{Abdo2011}. 
The low-frequency peak is certainly due to synchrotron emission from
relativistic electrons in the jet interacting 
with magnetic field; the high-frequency peak is probably due to inverse Compton
scattering of the same population of 
relativistic electrons with synchrotron low energy photons (Synchrotron Self
Compton model)\cite{Abdo2011, Tavecchio2001}. So we might expect to find an X-ray/$\gamma$-ray correlation. 
Mrk 421 shows variability at all frequencies; particularly at TeV energies 
Gaidos et al. \cite{Gaidos1996} measured a 
variability of $\sim15$ minutes. 

At radio frequencies this source clearly shows a one-sided jet structure aligned
at a small angle with respect 
to the line of sight \cite{Giroletti2006}.
In this work we present new VLBA (Very
Long Baseline Array)observations to study in detail the inner jet
structure on parsec scales. 
We can investigate the evolution of shocks that arise in the jet, 
with model-fitting techniques. In many works \cite{Piner1999, Piner2004} jet
components 
show only subluminal apparent motion, and this seems to be a common
characteristic of TeV Blazars. 
Thanks to accurate measurements of changes on parsec scales, by the VLBA, we can find
valid constraints on the geometry 
and kinematics of the jet. Despite several accurate studies on this source
\cite{Abdo2011},
details of physical processes responsible for the observed emission are still
poorly constrained. Because of its strong 
variability and broadband spectrum, multiwavelength  long term observations are
required for a good comprehension of 
emission mechanisms. 

This study is part of an ambitious multi-year and multi-instrument 
campaign, which also involves observations in sub-mm (SMA), optical/IR (GASP), UV/X-ray (Swift,
RXTE, MAXI), 
and $\gamma$ rays (Fermi-LAT, MAGIC, VERITAS). 
The aim of this observational effort is to try to shed light on fundamental questions such 
as the nature of radiating particles, the connection between radio and $\gamma$-ray emission, 
the location of emitting regions and the origin of the flux variability.
Very long baseline interferometry (VLBI) plays an important role in addressing these scientific questions because it is the only technique that can resolve (at least partially) the inner structure of the jet.
Therefore, cross-correlation studies of VLBA data with data from other energy ranges (in particular $\gamma$ rays) 
can provide us with important information about the structure of the jet and the location of the blazar emission.

\section{Observations}
We have new observations of Mrk 421 made throughout 2011 with the VLBA. 
The source was observed once per month, for a total of 12 epochs, at three 
frequencies: 15, 23.8 and 43 GHz.  In each epoch Mrk 421 has been observed for
nearly 
40 minutes. We also observed, with regular intervals, three other sources 
(J0854+2006, J1310+3220 \& J0927+3902) used as calibrators. In this paper we
present a preliminary 
analysis of the first seven epochs of observations (from January to July) at 15
and 23.8 GHz. 
For calibration and fringe-fitting we used the AIPS software package \cite{Greisen2003}, and for
map production 
we used standard self-calibration procedures included in the DIFMAP software
package \cite{Shepherd1997}. 
In some epochs one or more antennas did not work properly, because of technical
problems 
or bad weather conditions. For a complete report, see Table~\ref{observations}.

\section{Results}
\subsection{Images}
\begin{figure*}
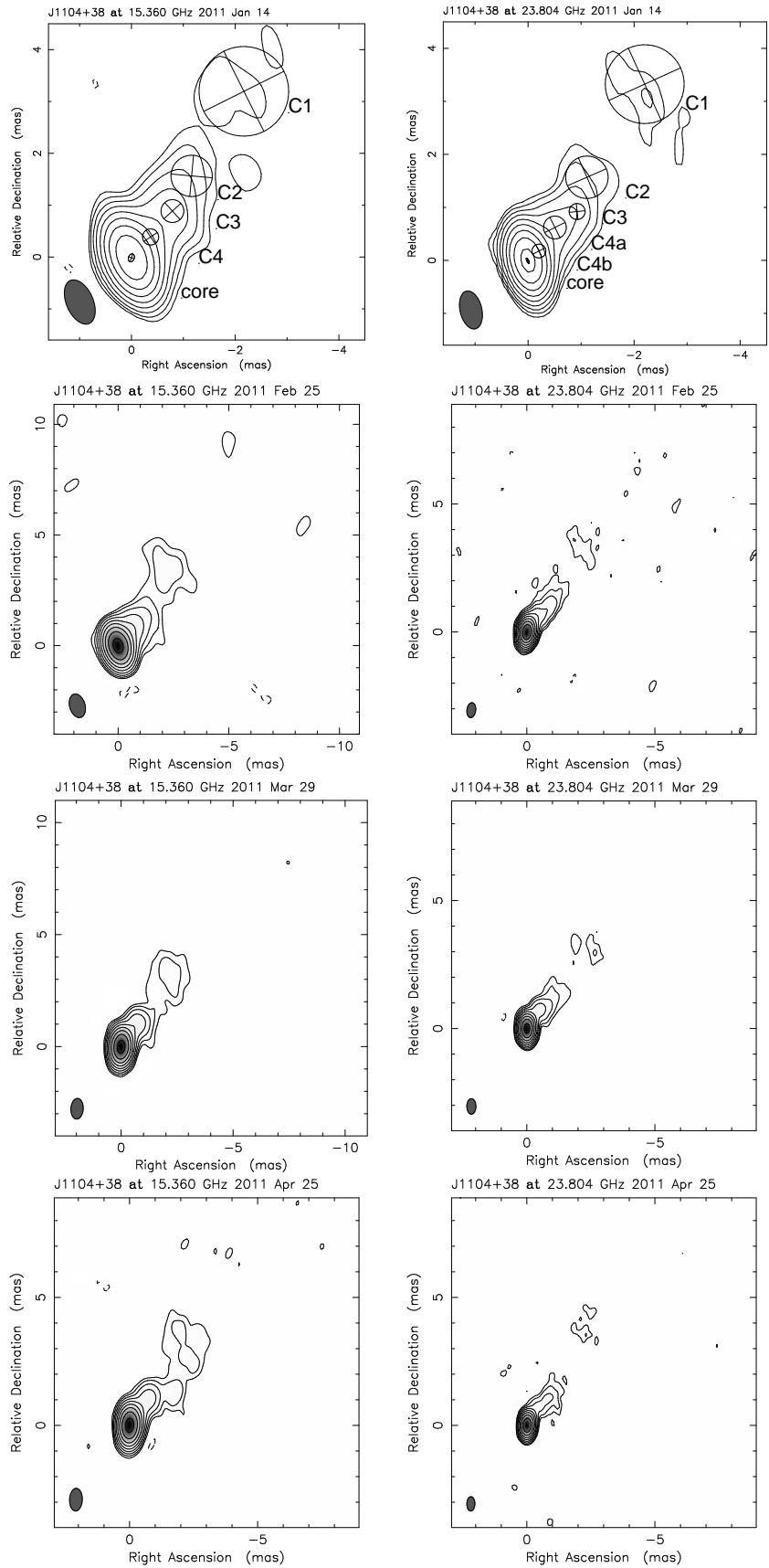

\centering
\includegraphics[angle=-90, clip, width=0.69\columnwidth]{gen15.ps} 
\includegraphics[angle=-90, clip, width=0.69\columnwidth]{gen24.ps} \\
\includegraphics[angle=-90, clip, width=0.693\columnwidth]{feb15.ps} 
\includegraphics[angle=-90, clip, width=0.693\columnwidth]{feb24.ps} 
\includegraphics[angle=-90, clip, width=0.693\columnwidth]{mar15.ps} 
\includegraphics[angle=-90, clip, width=0.693\columnwidth]{mar24.ps} 
\includegraphics[angle=-90, clip, width=0.693\columnwidth]{apr15.ps} 
\includegraphics[angle=-90, clip, width=0.693\columnwidth]{apr24.ps} 
\caption{Images of Mrk 421 from January to April. The left and right panels show
15 and 23.8 GHz images respectively. The restoring beam and the lowest contours
for each image are given in Table~\ref{observations}.
Levels are drawn at $(-1, 1, 2, 4...) \times$ the lowest contour given in Table~\ref{observations}.} 
\label{maps1}
\end{figure*}

\begin{figure*}
\centering
\includegraphics[width=0.7\columnwidth, angle=-90, clip]{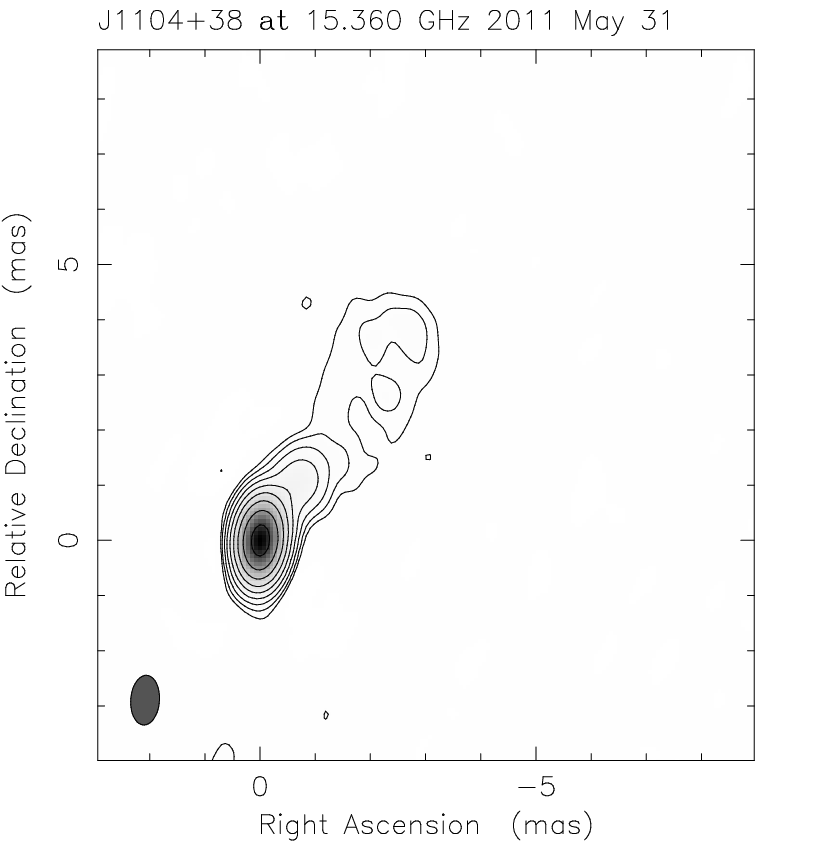} 
\includegraphics[width=0.7\columnwidth, angle=-90, clip]{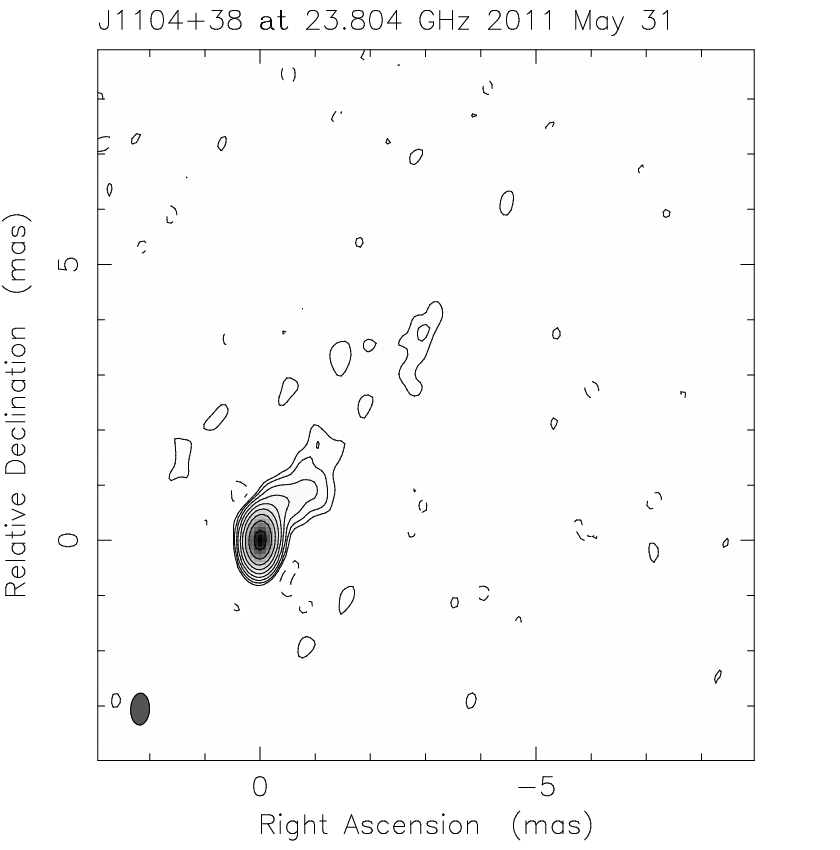} 
\includegraphics[width=0.7\columnwidth, angle=-90, clip]{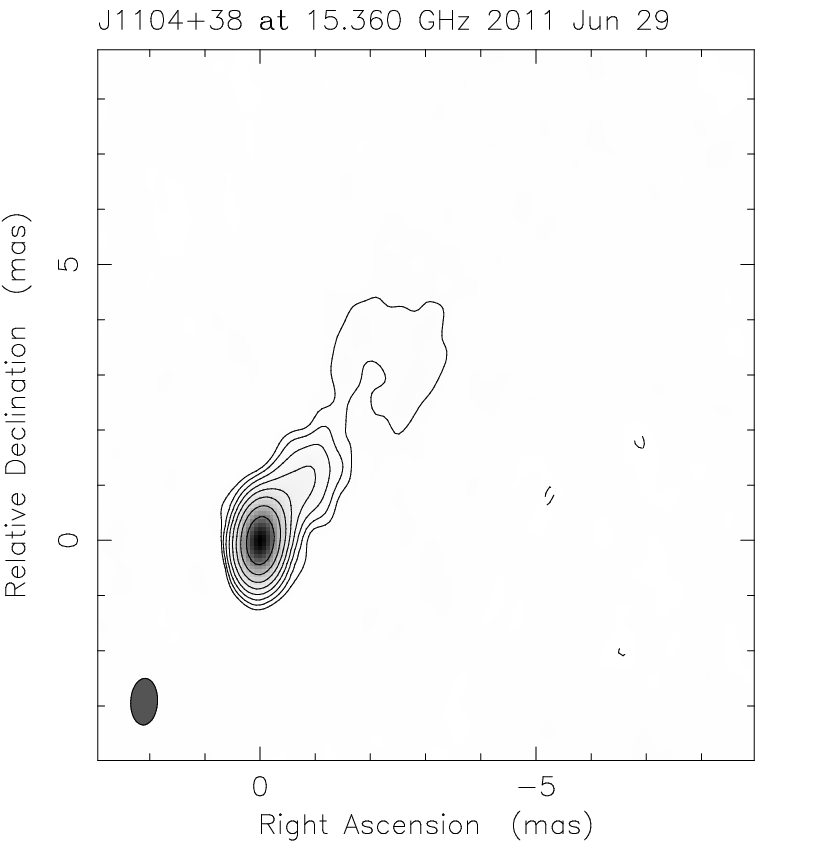} 
\includegraphics[width=0.7\columnwidth, angle=-90, clip]{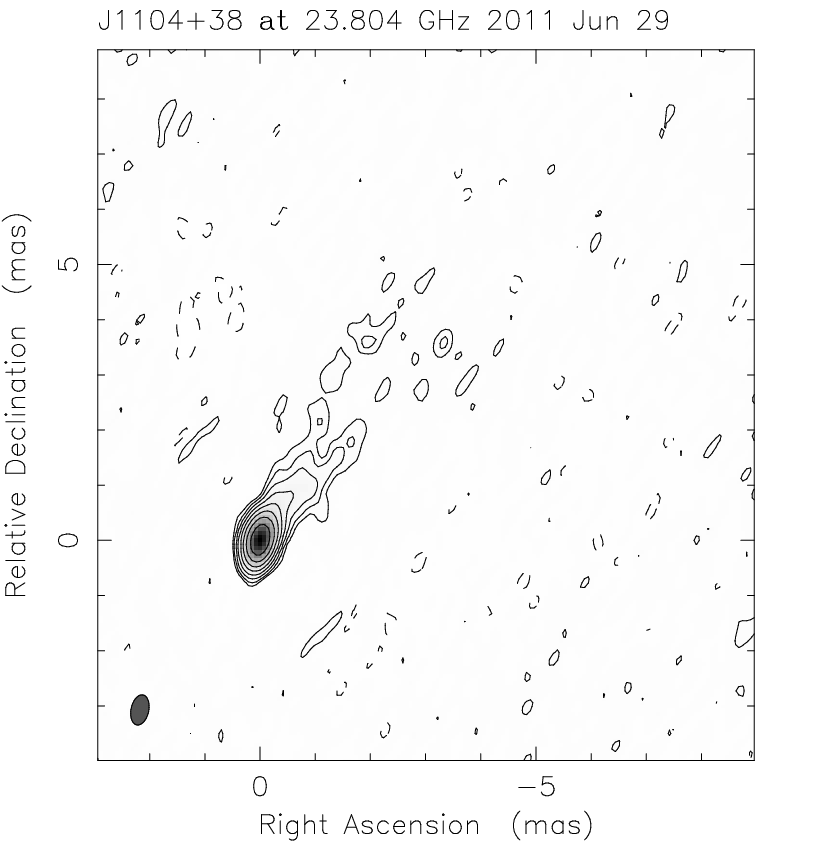} 
\includegraphics[width=0.7\columnwidth, angle=-90, clip]{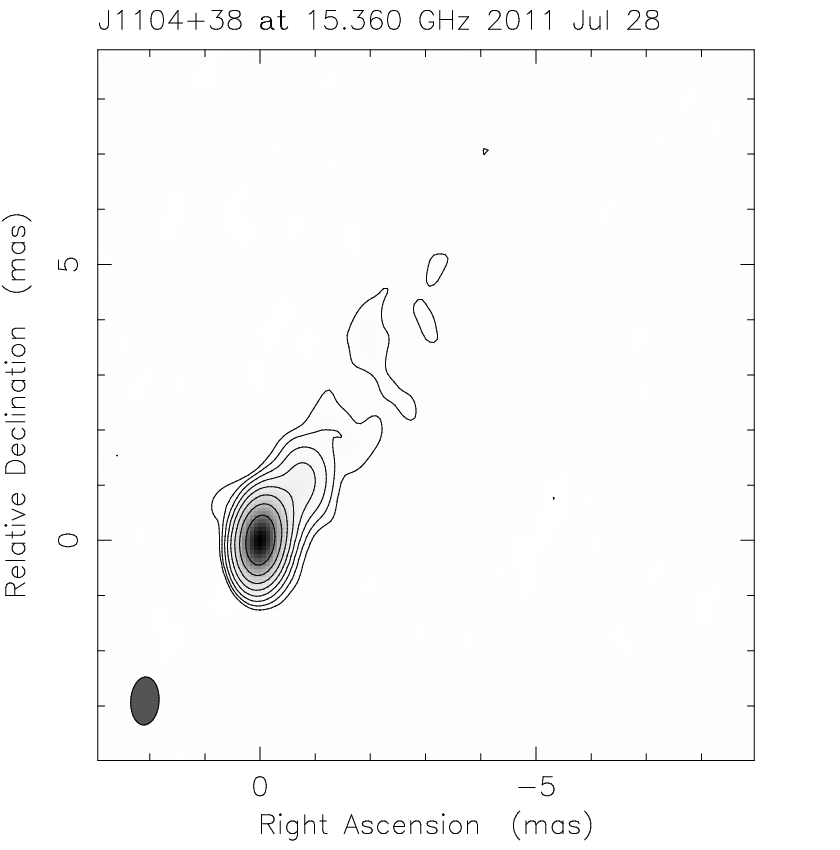} 
\includegraphics[width=0.7\columnwidth, angle=-90, clip]{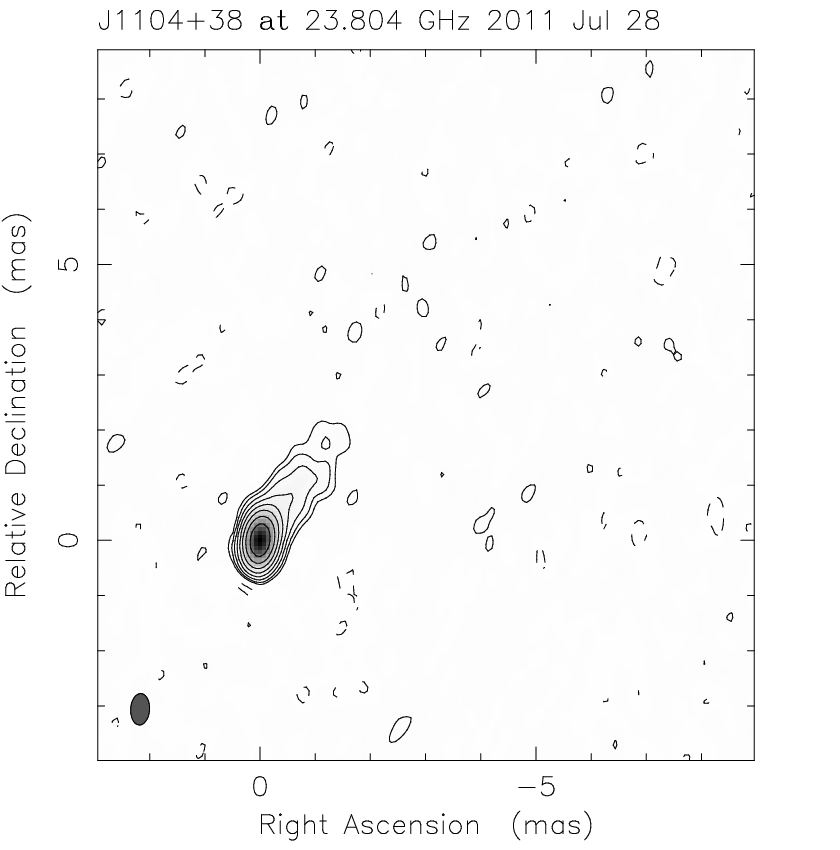} 
\caption{Images of Mrk 421 from May to July. The left and right panel show 15
and 23.8 GHz images respectively. The restoring beam and the lowest contours for
each image are given in Table~\ref{observations}.
Levels are drawn at $(-1, 1, 2, 4...) \times$ the lowest contour given in Table~\ref{observations}.} 
\label{maps2}
\end{figure*}

Images at 15 and 23.8 GHz for the first seven epochs created with DIFMAP 
are shown in Figs.\ref{maps1}  and \ref{maps2}. In all maps at these frequencies
Mrk 421 shows a well defined 
and well collimated jet structure (one-sided jet) emerging from a compact
nuclear 
region (core-dominated source). This is the typical structure of a BL Lac object
\cite{Giroletti2004a}. 
The jet extends for roughly 4.5 mas (2.67 pc), with a position angle slightly 
less than 45 deg in the northwest direction.

\subsection{Model fits}
For each epoch we used the model-fitting routine in DIFMAP to fit the
brightness 
distribution of the source in the u-v plane with elliptical or circular
Gaussian 
components. In this manner we can investigate in detail the inner jet structure
and its evolution. 
In all epochs at 15 GHz a good fit is obtained with five Gaussian components. At
23.8 GHz we have identified 
six components: thanks to the increased angular resolution, the second innermost
component (0.45 mas from the core, 
which we identify with the brightest, innermost and most compact component) is
resolved into 
two features (at 0.3 mas and 0.7 mas from the core). 

\begin{figure}
\centering
\includegraphics[clip,width=\columnwidth, height=0.9\columnwidth]{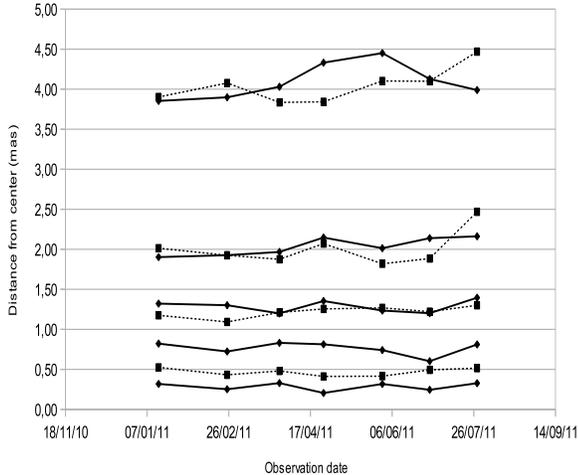} 
\caption{Results of model-fit analysis of the first seven epochs. Squares and
diamonds refer 
respectively to positions of Gaussian components at 15 and 23.8 GHz. Solid (23.8
GHz) and dashed (15 GHz) 
lines between data are only for guidance. Error bars (not shown) are comparable
with the positional 
scatter of each component. A proper treatment of the positional uncertainty, also
accounting for core opacity effects, 
will be given in a forthcoming publication.} 
\label{modelfit}
\end{figure}

At 15 GHz we label components with C1, C2, C3, C4, starting from the
outermost (C1) to the innermost (C4). 
At 23.8 GHz the C4 component is split into C4a and C4b. 
Overall, the components extend up to a region of about 5 mas. 
In this way, with a limited number of components, it is possible 
to analyze proper motions at various times and flux levels. 
From Fig.\ref{modelfit} we can clearly see 
that data occupy well-defined areas in the plane, and this behavior supports the
identification of individual 
components across epochs. However this choice is not unique, so it is
useful to confirm with the analysis of flux variations discussed in the next section. We will
investigate all these 
aspects in more detail in a further analysis of data at 43 GHz.

\subsection{Flux density  variations}
Using the modelfit it is also possible to analyze the temporal evolution of
the flux 
for each component of the source. The brightest component is the one describing the nuclear
region 
(core-dominated source); at 15 GHz it has a value around 350 mJy, which
decreases along 
the jet as it moves away from the core, to values ​​around 10 mJy. From a comparison of the flux density of each component at the various epochs it 
emerges that there is no significant variation in flux densities for both the
core and 
for other components; the flux value for each component remains roughly constant
at various 
times within the uncertainties calculated. Small variations may be artifacts
brought about by our fitting procedures: e.g. 
the flux density of inner components may be underestimated in some cases,
because part of it 
was incorporated into the core component flux. This is a remarkable result because
it 
allows the identification of components based on flux density values, and it 
confirms our choice of components based on their positions. A comparison
to the lightcurves at higher 
frequencies, as well as the discussion of the spectral index structure, will be
the subject of a future paper.

\subsection{Apparent speeds}
From our modelfit we infer low or no displacement for the jet components. To quantify this value we have determined speeds for each component through
linear 
fits to the separation of the individual features from the core in different
epochs. For the 
three outer components (C1, C2 and C3), we used combined data at 15 and at 23.8
GHz, since the positions of each component at the two frequencies are consistent within the error bars; 
for the two inner components (C4a and C4b) we used only data at 23.8 GHz.
Results are shown in Table~\ref{speeds}.

We find low values for the apparent speed, which agrees with previous studies \cite{Piner2005}: 
the two innermost
components (C4a and C4b) are consistent with being stationary; C2 and C3 have
low-significance (1-2$\sigma$) 
subluminal motion, and only the outermost component C1 shows superluminal
motion, although only at 2.5$\sigma$ 
confidence level.
We will give a possible interpretation in the discussion section. It is
important to emphasize that these are 
only partial results from a preliminary analysis; we can not exclude that they will change 
significantly in the conclusive analysis of the entire data set. 
We will give more accurate conclusions in a forthcoming paper.

\begin{table}[t]
\begin{center}
\footnotesize
\begin{tabular}{lcc}
\hline
Component & Apparent speed & $\beta_{\rm app}$ \\
& (mas/month) &  \\
\hline
\hline
C1   &$0.06\pm0.02$ & $1.3\pm0.5$ \\
C2   &$0.04\pm0.02$ & $0.9\pm0.4$ \\
C3  &$0.01\pm0.01$ & $0.3\pm0.2$ \\
C4a  &$-0.01\pm0.02$ & $-0.3\pm0.4$ \\
C4b   &$0.00\pm0.01$ & $0.0\pm0.2$  \\
\hline
\end{tabular}
\caption{Apparent speeds from linear fit analysis.\label{speeds}}
\end{center}
\end{table}

\section{Discussion and conclusions}
The trend of the measured velocities can be interpreted in different ways. If
these apparent speeds, 
shown in Table~\ref{speeds}, represent the bulk apparent speed of the plasma in the jet,
we can infer that some 
mechanism involving an acceleration acts in the outer region of the jet,
resulting in a higher speed 
for the outermost component, e.g. C1. 
Alternatively we can invoke the presence of a tranverse velocity structure along
the jet axis. 
This structure consists of two components: a fast inner \textit{spine} and a
slower outer \textit{layer}. 
Speeds are obtained depending on whether we are measuring the speed of the spine
or the layer. In this way 
speeds reported in Table~\ref{speeds} are only pattern speeds, and we suggest that various
blobs of plasma (components) 
that represent the jet's structure found by us do not represent moving structures. Therefore, 
measured velocities would not be intrinsically linked to the jet's bulk velocity.

Other variations of this model have been proposed
\cite{Giroletti2004b,Ghisellini2005}, and we expect to constrain the
variability and the details of these models with the complete analysis of the
whole dataset, giving our own interpretation.
In any case, it is difficult to reconcile slow apparent speeds in the inner
region of the jet with the high value for Doppler factor 
($\delta$)\footnote{$\delta=1/\gamma(1-\beta cos\theta)$, where $\gamma$ is the
bulk Lorentz factor, $\theta$ is the angle of view and $\beta=v/c$.} 
required from TeV variability. With the $\delta$ dependence on both $\beta$ and
$\theta$, only 
very small viewing angles ($\theta \sim 1^\circ$) are compatible with the measured
apparent speeds. 
While it is true that such a value is not particularly surprising for a blazar,
we note that similar (or even smaller) values for $\beta_{app}$ are found for all the $\sim$10 TeV blazars that have
proper motions reported in the 
literature \cite{Piner2008, Piner2010}.
When angles of a small fraction of a degree are required for many sources, then
the 
statistics are untenable, and the ``Doppler crisis'' for the TeV blazars can not
be explained solely on the basis of small 
angles to the line of sight (see also \cite{Lyutikov2010}). 

These findings might
also be ascribed to the lack of high contrast
features in the model fit.  Most likely, the low apparent speed found implies
that proper motion in this case does not give any information on the jet bulk
velocity that, according to other evidence (jet sidedness, core dominance, and high energy emission), 
requires a relativistic jet velocity.
Indeed, we plan 
to further constrain the jet velocity by means of a study of the jet sidedness
ratio. We will present the entire 
analysis of constraints for these physical parameters in a forthcoming paper.

In summary, the preliminary analysis presented here has confirmed with high clarity
the predictions made on the basis of 
knowledge so far achieved for TeV blazars. 
However, there is still much to be understood and we expect to obtain more
significant results from the analysis of the 
entire dataset, particulary from 43 GHz data. 
Finally we could combine our data with those of other works \cite{Piner2005} by
increasing the temporal coverage 
of observations to obtain meaningful results over a longer time frame.

\bigskip 
\begin{acknowledgments}
This work is based on observations obtained through the BG207 VLBA project. The
National Radio 
Astronomy Observatory is a facility of the National Science Foundation operated
under cooperative 
agreement by Associated Universities, Inc.
\end{acknowledgments}

\bigskip 

\end{document}